\documentclass[twocolumn,showpacs,preprintnumbers,amsmath,amssymb,aps,10pt,floatfix]{revtex4-1}
\usepackage{graphicx}
\usepackage{epstopdf}
\usepackage{dcolumn}
\usepackage{bm}
\usepackage{color}

\begin{document}

\title{Inhomogeneous nuclear spin polarization induced by helicity-modulated optical excitation of fluorine-bound electron spins in ZnSe}

\author{F.~Heisterkamp$^{1}$, A.~Greilich$^{1}$, E.~A.~Zhukov$^{1}$, E.~Kirstein$^{1}$, T.~Kazimierczuk$^{1}$, V.~L.~Korenev$^{1,2}$, I.~A. Yugova$^{3}$, D.~R.~Yakovlev$^{1,2}$, A.~Pawlis,$^{4}$ and M.~Bayer$^{1,2}$}

\affiliation{$^1$ Experimentelle Physik 2, Technische Universit\"at
Dortmund, 44221 Dortmund, Germany}

\affiliation{$^2$ Ioffe Institute, Russian Academy of Sciences, 194021 St.~Petersburg, Russia}

\affiliation{$^3$ Physical Faculty of St.~Petersburg State
University, 198504 St.~Petersburg, Russia}

\affiliation{$^4$ Peter Gr\"unberg Institute (PGI-9), Forschungszentrum J\"ulich, 52425 J\"ulich, Germany}

\date{\today}

\begin{abstract}
Optically-induced nuclear spin polarization in a fluorine-doped ZnSe
epilayer is studied by time-resolved Kerr rotation using resonant
excitation of donor-bound excitons. Excitation with
helicity-modulated laser pulses results in a transverse nuclear spin
polarization, which is detected  as a change of the Larmor
precession frequency of the donor-bound electron spins. The
frequency shift in dependence on the transverse magnetic field
exhibits a pronounced dispersion-like shape with resonances at the
fields of nuclear magnetic resonance of the constituent zinc and
selenium isotopes. It is studied as a function of external
parameters, particularly of constant and radio frequency external
magnetic fields. The width of the resonance and its shape indicate a
strong spatial inhomogeneity of the nuclear spin polarization in the
vicinity of a fluorine donor. A mechanism of optically-induced
nuclear spin polarization is suggested based on the concept of
resonant nuclear spin cooling driven by the inhomogeneous Knight
field of the donor-bound electron.
\end{abstract}


\maketitle

\section{Introduction}\label{sec:intro}

The hyperfine interaction between electron and nuclear spins in
semiconductor structures has been of particular interest over many
years~\cite{OptOR,Kalevich}. Lately it was intensively driven by
the intention to extend the electron spin coherence time and the
idea to employ the nuclear spins as quantum bits
(qubits)~\cite{reviewNuc}. Using optical excitation with circularly
polarized light angular momentum can be transferred via the electron
system to the nuclei. The polarized nuclei, in turn, act back on the
electrons as an effective magnetic field (the Overhauser field),
causing a splitting of the electron spin states~\cite{OptOR}.
Without an external magnetic field, the hyperfine interaction with
the nuclear spins is the main source of dephasing for localized
electron spins in which fluctuations in the nuclear spin
polarization induce fluctuations in the electron spin
splittings~\cite{Khaetskii02,Merkulov02,Dzhi02}.

The nuclei-induced electron spin splitting can reach values of
100\,$\mu$eV for high nuclear spin polarization. It can be detected
spectrally as a splitting of the emission line, if the linewidth of
the photoluminescence (PL) is sufficiently narrow, as it is the case
for single emitters like quantum dots (QDs) or impurity centers. A
relatively high nuclear spin polarization can be created using the
method of optical pumping in longitudinal magnetic field (parallel
to the optical axis and orthogonal to the sample
surface)~\cite{Snow96,Gammon01,Braun06,Chekhovich13}. In the case of
inhomogeneously broadened optical transitions, as it is the case in
emitter ensembles, optical methods do not provide sufficient
spectral resolution so that alternative approaches should be used.
The nuclear spin polarization can be indirectly measured by its
influence on the electron spin polarization in a transverse magnetic
field, the Hanle effect~\cite{Hanle}. In this case the Overhauser
field modifies the measured degree of the circular PL polarization
as a function of magnetic field applied in the direction transverse
to the optical
excitation~\cite{Kalevich,Paget77,Cherbunin09,Auer09,Krebs10,Flisinski10}.
If one additionally applies a polarization modulation of the
excitation at a frequency corresponding to the nuclear magnetic
resonance of one of the constituent isotopes, a nuclear spin
polarization in the direction transverse to the optical excitation
can be achieved~\cite{Kalevich1980,Kalevich1995,Cherbunin11}.
However, such measurements are limited to the field ranges within
the width of the Hanle curve, which is determined by the spin
dephasing time of the carriers, their $g$ factor values and the
nuclear spin polarization.

In the recent publication of Zhukov \textit{et al.}~\cite{Zhu2014}
the authors extended the magnetic field range by implementing a
novel experimental technique to measure the optically-induced
nuclear magnetic resonances by assessing the electron spin coherence
in the resonant spin amplification (RSA) regime. For that purpose
time-resolved pump-probe Kerr rotation (TRKR) was
used~\cite{Kikkawa98,Yugova12}. Information about the Overhauser
field that changes the electron spin splitting was thereby
transferred from the spectral domain into the temporal domain. In
the present paper we extend the investigation of dynamic nuclear
spin polarization (DNP) by applying the developed experimental
approach to ZnSe doped with fluorine donors. We observe changes of
the electron Larmor precession in a broad range of transverse
magnetic fields, which depends on the pump power. The induced shifts
of the Larmor precession change sign at the fields of the nuclear
magnetic resonances (NMRs). A theoretical approach to analyze the
mechanism of dynamical nuclear spin polarization for our
experimental conditions is developed.

Fluorine donors in ZnSe (further ZnSe:F) and the corresponding
donor-bound electron spins are currently of substantial interest
because of several recent key achievements towards solid-state
quantum devices: Sources of indistinguishable single
photons~\cite{Sanaka09} as well as entangled
photon-pairs~\cite{SanakaNL12} and optically controllable
electron-spin qubits~\cite{DeGreve10,Sanaka12,SleiterNL13} have been
demonstrated so far. The spin coherence of the donor-bound electron
is generally limited by the non-zero nuclear spin background in the
host crystal~\cite{Merkulov02}. However, in ZnSe isotopic
purification can be applied to deplete the remaining low amount of
non-zero nuclear spins. Apart from that, fluorine has a natural
100\% abundance of spin 1/2 nuclei, which might be considered as a
nuclear spin qubit coupled to a single electron spin qubit via the
hyperfine interaction. These aspects make the ZnSe:F system
particularly attractive for the investigation of electron and
nuclear spin related features.

\section{Experimental details} \label{sec:sample}

The sample under study is a homogeneously fluorine-doped,
70-nm-thick ZnSe:F epilayer grown by molecular-beam epitaxy on a
$(001)$-oriented GaAs substrate. The epilayer is separated from the
substrate by a 20-nm-thick
$\text{Zn}_{0.85}\text{Mg}_{0.15}\text{Se}$ barrier layer to prevent
carrier diffusion into the substrate. The barrier in turn is grown on top of a thin ZnSe buffer layer to reduce the defect density at the III-V/II-VI heterointerface. The concentration of fluorine
donors in the ZnSe:F epilayer is about $10^{18}$\,cm$^{-3}$, so that
the distance between the neighboring donors is larger than the Bohr
radius of the donor-bound electrons~\footnote{Similar effects were observed for fluorine concentrations down to $10^{15}$\,cm$^{-3}$.}. The sample was placed in an optical cryostat with a superconducting split-coil magnet. The sample temperature was fixed at $T=1.8$~K in all measurements.

\begin{figure}[t]
\includegraphics[width=\columnwidth]{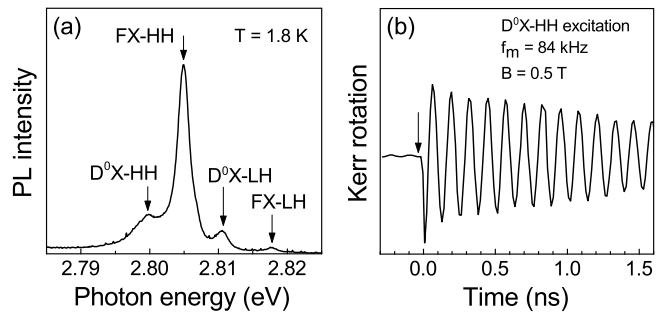}
\caption{(a) Normalized PL spectrum measured at $B=0$\,T and
$T=1.8$\,K. (b) Kerr rotation signal measured for resonant
D$^{0}$X-HH excitation at 2.800\,eV. $B=0.5$\,T. The arrow marks the
negative time delay of the probe with respect to the pump at which
the RSA signal is measured.} \label{fig:PL}
\end{figure}

Figure~\ref{fig:PL}(a) shows the normalized PL spectrum measured for
continuous-wave (cw) excitation with a photon energy of 3.05\,eV.
The PL is detected with a Si-based charge-coupled-device camera
attached to a 0.5\,m spectrometer. In the characteristic peak
pattern each feature can be assigned to different exciton
complexes~\cite{PL_diss}. The labels in the figure mark the
following optical transitions: FX refers to the free exciton and
D$^0$X to the donor-bound exciton containing heavy holes (HH) or
light holes (LH), respectively.

To obtain insight into the electron spin dynamics we use the TRKR
technique. The electron spin coherence is generated by
circularly-polarized pump pulses of 1.5\,ps duration (spectral width
of about 1\,meV) generated by a mode-locked Ti:Sa laser operating at
a repetition frequency of 75.75\,MHz (repetition period
$T_{\text{R}}=13.2$\,ns). The laser frequency is doubled by a BBO
(beta barium borate) crystal to convert the Ti:Sa range of photon
energies from about $1.25-1.7$\,eV to $2.5-3.4$\,eV. After
excitation of the sample along the growth axis \textbf{z} with the
circularly-polarized pump pulses, the reflection of the
linearly-polarized probe pulses is analyzed with respect to the
angle of polarization rotation as a function of delay between the
pump and probe pulses. The pump helicity is modulated between
$\sigma^+$ and $\sigma^-$ circular polarization by an
electro-optical modulator (EOM) using modulation frequencies
$f_\text{m}$ in the range $20-500$\,kHz. The probe beam is kept
unmodulated. The Kerr rotation (KR) angle is measured with a
balanced detector connected to a lock-in amplifier.

Figure~\ref{fig:PL}(b) shows a TRKR signal measured at a magnetic
field of $B = 0.5$\,T applied perpendicular to the growth axis
\textbf{z} (Voigt geometry). Both pump and probe have the same
photon energy (degenerate pump-probe scheme) and are resonant with
the D$^0$X-HH transition at 2.800\,eV. The pump power is $P=1.1$\,mW
and the probe power 0.5\,mW, both having a spot size of about
300\,$\mu$m on the sample. The oscillations in the TRKR signal
correspond to the Larmor spin precession of the donor-bound electron
with a $g$ factor of $g_\texttt{e} =1.13 \pm
0.02$. The oscillating signal at negative
pump-probe delays indicates that the spin coherence is maintained up
to the arrival of the next pump pulse, i.e. the electron spin
dephasing time $T_2^*$ is close to the laser repetition period
$T_{\text{R}}$. Due to the long spin dephasing time the action of
the succeeding pump pulses on the remaining spin polarization
depends on the electron $g$ factor and the magnetic field strength.
In this case, the resonant spin amplification regime can be
used~\cite{Kikkawa98,Yugova12} for studying spin coherence. To that
end we fix the probe at a slightly negative time delay of $\Delta t=
-20\,$ps with respect to the pump pulse arrival moment (see arrow in
Fig.~\ref{fig:PL}(b)) and detect the KR signal while scanning the
transverse magnetic field. Further, we conducted experiments where
an additional oscillating magnetic field (radio frequency or
RF-field) $\mathbf{B}_{\mathrm{RF}}=(0,0,B_{\mathrm{RF},z})$ is
applied along the $z$ axis using a small coil placed close to the
sample surface.

\section{Experimental results} \label{sec:exp}

\subsection{Observation of the nuclear spin polarization}\label{subsec:experiment}

\begin{figure}[t]
\includegraphics[width=\columnwidth]{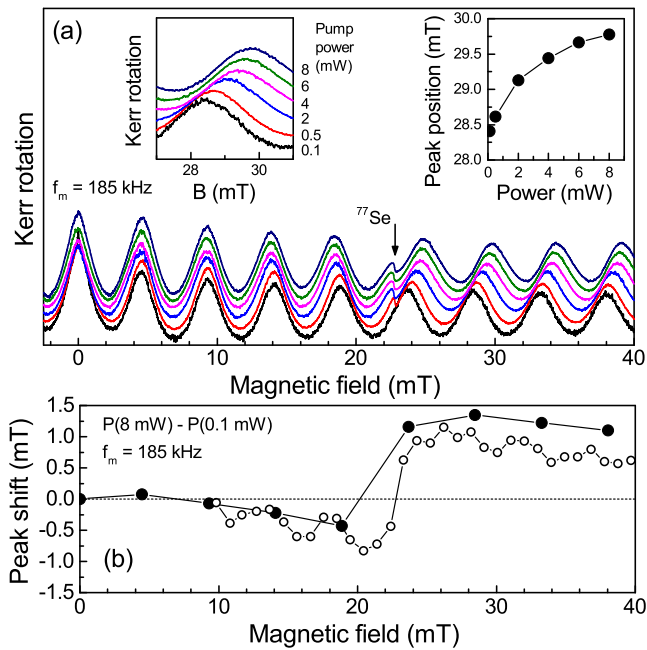}
\caption{(Color online) (a) Normalized RSA signals measured at
different pump powers and $f_\text{m}=185$\,kHz. The NMR of
$^{77}$Se at $B=22.76$\,mT is marked by the arrow. The left inset
shows a close up of the RSA peaks around 29\,mT. The position of the
peak maximum is plotted as function of the pump power in the right inset.
$T=1.8$\,K. (b) Black solid circles show how the difference (shift)
between the peak positions at 8\,mW and at 0.1\,mW from panel (a)
changes in dependence on the magnetic field. Open circles
demonstrate the induced shift measured directly by time-resolved KR
(see text). Lines are guides to the eye.} \label{fig:main}
\end{figure}

Figure~\ref{fig:main}(a) shows a set of RSA curves of normalized
amplitude for an EOM modulation frequency of $f_\text{m} =
185$\,kHz, measured for different pump powers. The RSA curves have
the typical shape of periodic peaks as function of the magnetic
field strength. The dependencies of the RSA peak distance and width
on magnetic field are determined by the electron $g$ factor
$g_\texttt{e}$, the $g_\texttt{e}$ spread $\Delta g$, the spin
dephasing time $T_2^*$ and the repetition period of the laser pulses
$T_{\text{R}}$~\cite{Kikkawa98,Yugova12}. Further optical properties
of this sample and information on the electron spin dephasing and
relaxation mechanisms can be found in Refs.~\cite{Greilich12,
Heist2015}.

Let us turn now to the experimental results on nuclear effects
induced and detected via the spin dynamics of the resident
donor-bound electrons. As mentioned above, the sample is excited by
the helicity modulated pump. Under these conditions nuclear spin
polarization in the system is usually suppressed, as the nuclei
cannot follow the oscillating electron spin polarization fast enough
due to the long nuclear spin relaxation time. However, nuclear spin
polarization can be observed for finite magnetic fields in close
vicinity to the NMR frequencies (corresponding to the helicity
modulation frequency) of the constituent isotopes. Here the nuclear
spin system becomes polarized through its cooling in the magnetic
field which results from the electron spin polarization (Knight
field)~\cite{Knight1949, Zhu2014}. The induced nuclear spin
polarization appears in RSA curves as an additional amplitude
modulation at the NMR fields, see in Fig.~\ref{fig:main}(a) the
narrow resonance for the $^{77}$Se isotope at 22.76\,mT measured for
$f_\text{m} = 185$\,kHz at pump powers above $0.5$\,mW.

If we consider the behavior of the RSA peaks at different excitation
powers we observe the following behavior: Fig.~\ref{fig:main}(a)
displays several RSA curves taken at pump powers varied from
$P=0.1$\,mW (bottom spectrum) up to 8\,mW (top spectrum). The probe
power is kept in the range of $0.5$\,mW for all measurements.
The signal amplitudes are normalized and the curves are displaced
vertically relative to each other, to simplify comparison of the
peak positions. The increase of the pump power leads to a shift of
the RSA peaks, which reflects a change of the electron precession
frequency. The left inset in Fig.~\ref{fig:main}(a) provides a
closeup of a single peak to highlight this shift. The peak positions
for each curve are given in the right inset, which evidences a
saturation effect at high pump powers.

One can see from Fig.~\ref{fig:main}(a), that the direction and the
magnitude of the shift depend on the RSA peak position relative to
the NMR field and also on the pump power. The peaks located at lower
fields relative to the resonance exhibit a shift towards lower
magnetic fields for increasing pump power, while the peaks at higher
fields than the resonance are shifted towards higher magnetic field.
The shift itself can be explained by an additional induced magnetic
field acting in addition to the external magnetic field on the
electron spins. The shift of RSA peaks to lower magnetic fields
indicates an additional magnetic field pointing in the same
direction as the external field. Vice versa, the shift to higher
fields indicates opposite orientations of the induced and external
fields.

To demonstrate how the peaks shift across the whole range of
external magnetic fields, we plot the difference of the peak
position taken at maximal (saturated, $P=8$\,mW) and minimal used
pump powers (0.1\,mW) as a function of magnetic field. The black
solid circles in Fig.~\ref{fig:main}(b) show this dependence. The
circles are placed at the fields of the RSA peaks measured at
minimal pump power. The induced RSA peak shift (vertical axis in
Fig.~\ref{fig:main}(b)) directly corresponds to the opposite
strength of the induced nuclear field.

\begin{figure}[t]
\includegraphics[width=\columnwidth]{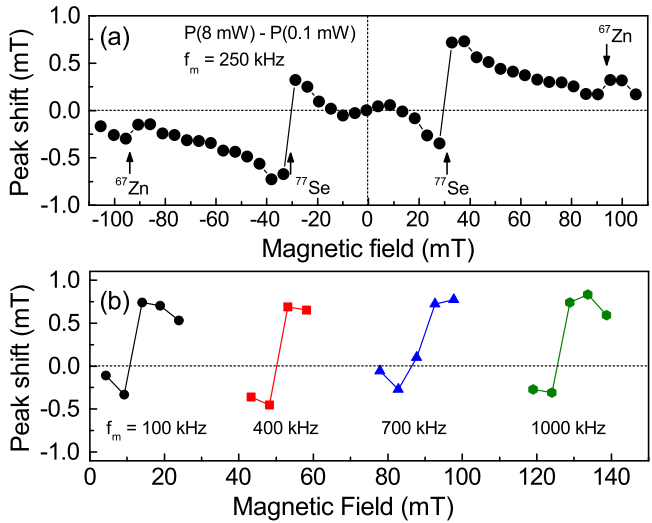}
\caption{(Color online) (a) Induced RSA shift measured at
$f_\text{m}=250$\,kHz in a broader range of magnetic fields compared
to Fig.~\ref{fig:main}. Therefore the NMR of the $^{67}$Zn isotope
is also observed. (b) Shift of the RSA peaks measured at different
modulation frequencies $f_\text{m}$.} \label{fig:mod}
\end{figure}

Figure~\ref{fig:mod}(a) shows the pump induced shift over a broader
range of magnetic field, measured at a modulation frequency of
$f_\text{m}=250$\,kHz. Here the NMR of the $^{77}$Se and $^{67}$Zn
isotopes are observed at $|B|=30.75$\,mT and $93.75$\,mT,
respectively. Figure~\ref{fig:mod}(b) shows in addition the RSA
shifts in the vicinity of the $^{77}$Se NMR measured for various
$f_\text{m}$ up to $1$\,MHz. The amplitude and the shape of the
shifts remain unchanged in this field range.

The dependencies in Figs.~\ref{fig:main}(b) and~\ref{fig:mod}
exhibit several peculiarities:
\begin{itemize}
\item[(i)] They show a characteristic dispersive shape around the NMRs of
the constituent isotopes. In Fig.~\ref{fig:mod}(a) one sees these
resonance-like dispersive features for both the selenium and zinc
isotopes. In the following we will call them resonances for
simplicity.
\item[(ii)] The resonances occur at magnetic fields, which are much larger
than the width of the Hanle curve. Here the Hanle-curve width is
given by the width of the RSA peak around zero magnetic field and is
about 2\,mT (full width half maximum - FWHM), see also
Ref.~\cite{Zhu2014}.
\item[(iii)]
The resonances are extended over a quite broad magnetic field range
as result of their slowly decaying dispersive tails.
\end{itemize}

For a deeper understanding, the open circles in
Fig.~\ref{fig:main}(b) show the results of additional
measurements, where the induced nuclear fields are not only given at
the RSA peak positions, but are also determined in between the peaks
to obtain a higher resolution on magnetic field. To provide these
data we determine the electron precession frequency from a
measurement of the KR signal as a function of the time delay between
pump and probe pulses (see Fig.~\ref{fig:PL}(b)). We perform the
measurements for transverse magnetic fields varied from 10\,mT up to
40\,mT using a 1\,mT incremental step. Each data point represents
the difference of the electron Larmor frequencies measured at 8\,mW
and 0.1\,mW pump power. Using the electron $g$ factor
$g_\texttt{e}=1.13$, one can convert this Larmor frequency
difference to a magnetic field, which corresponds to the RSA peak
shift, so that we can present it on the same scale as the RSA
measurements.

One sees that the experimental data given by the open circles in
Fig.~\ref{fig:main}(b) follow closely the solid circles. Both curves
show a slight vertical shift to positive values. The FWHM of the
broad dispersive resonance is on the order of 40\,mT. The high
resolution of the open circles data is particularly advantageous for
the transition region slightly above 20~mT where the sign reversal
of the RSA peak shift takes place. This sign reversal occurs in a
field range of $1-2$\,mT only. Note that the long tails and the
sharp transition cannot be explained by a simple dispersive curve
with a large homogeneous linewidth, which will be addressed in more
detail by our model considerations in Sec.~\ref{subsec:theory}.
Additionally, one can see that the open circles data exhibit
oscillations with a period equal to the period of the RSA peaks.
These oscillations could not be resolved in the data set given by
the closed circles, as there the measurements were performed only at
the RSA peak positions.

\begin{figure}[t]
\includegraphics[width=\columnwidth]{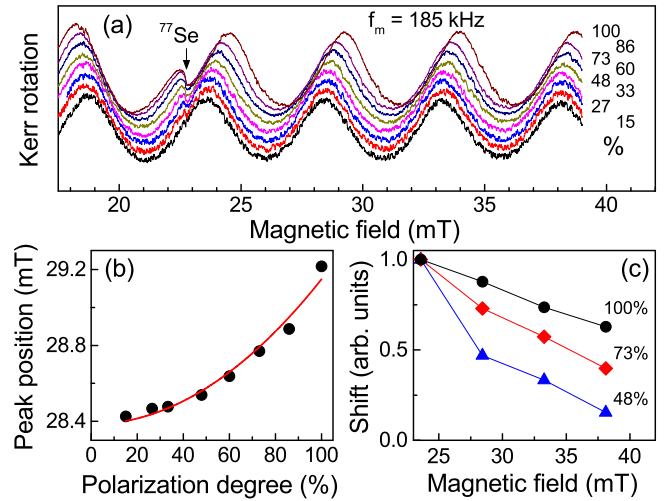}
\caption{(Color online) (a) Normalized RSA spectra for different
circular polarization degrees of the pump. The spectra are shifted
vertically relative to each other for clarity. $T=1.8$\,K,
$P=8$\,mW. (b) Position of the RSA peak around 28\,mT field strength
in dependence on the degree of circular polarization. Red line is a
parabolic fit to the data demonstrating the quadratic dependence of
the nuclear spin polarization on the circular polarization. (c)
Relative peak shifts for different circular polarization degrees in
dependence on the magnetic field. The data are normalized to the
shift of the peak at 24\,mT.} \label{fig:SP}
\end{figure}

\subsection{Electron spin polarization} \label{subsec:spinpol}

Before we turn our attention to the investigation of the
peculiarities listed in the previous section, we check how the
induced nuclear spin polarization depends on the electron spin
polarization. Figure~\ref{fig:SP}(a) shows the measured RSA signals
for several degrees of pump circular polarization at
$f_\text{m}=185$\,kHz and $P=8$\,mW. The signals are normalized in
amplitude to simplify comparison of the peak shift. One clearly sees
for the RSA peak close to the NMR, that the pump-induced shift
decreases for lower circular polarization degrees. This confirms
that the nuclear spin polarization is induced by spin-oriented
electrons. Figure~\ref{fig:SP}(b) shows an example of the peak shift
dependence on the polarization degree around 28\,mT. The red line
represents a parabolic fit to the data, demonstrating the quadratic
dependence of the nuclear spin polarization on the degree of
circular polarization.

Additionally, Fig.~\ref{fig:SP}(c) shows how the peak-shift depends
on the magnetic field for three different degrees of polarization.
Obviously the width of the resonance is proportional to the induced
nuclear spin polarization and decreases for lower polarization
degrees of the electron spin. Here the shifts are given relative to
the peak position for $15$\% polarization degree. The shift
amplitudes are then normalized to the shift of the peak at about
$24$\,mT. This allows us to clearly resolved the accelerated decay
with magnetic field for lower degrees of the nuclear spin
polarization.

For a complete understanding of the nature of the nuclear spin
polarization produced by the polarized electron spins, information
about the average electron spin polarization $\mathbf{S}$ is
important~\cite{Zhu2014}. To decide which spin component is
responsible for the nuclear spin polarization we should provide a
tomographic measurement for all three of them: $S_x$, $S_y$ and
$S_z$.

\subsubsection{$S_z$-component: Time-resolved Kerr rotation}\label{subsubsec:Sz}

\begin{figure}[t]
\includegraphics[width=\columnwidth]{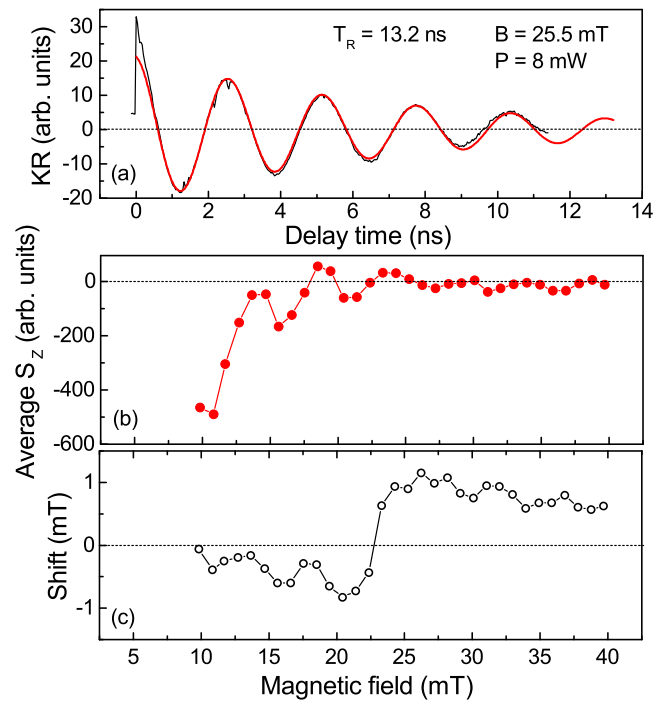}
\caption{(Color online) (a) Example of a time resolved pump-probe
Kerr signal measured at $B=25.5$\,mT (black curve). Red curve is a
fit to the data with an exponentially decaying cosine function
recorded over the full range of delays of $T_\text{R}=13.2$\,ns
between two pump pulses. (b) Average electron spin polarization
along the $z$ axis resulting from the fitted curves as a function of
magnetic field. (c) Reproduction of the data from
Fig.~\ref{fig:main}(b) for simplified comparison with the data in
panel (b).} \label{fig:Sz}
\end{figure}

In our experiment the average $S_z$ component, which is created
along the optical excitation axis can be evaluated directly from the
experimental data by integrating the KR signal over the whole time
period between the pump pulses. This averaging is expected to result
in a finite value of $S_z$ in magnetic fields for which the Larmor
precession period, $T_\texttt{L}$, is longer than or comparable with
the spin dephasing time $T^*_2$. this relation is not fulfilled in
strong magnetic fields, where $T_\texttt{L} \ll T^*_2$, $S_z \approx
0$, but in the field range studied here it is perfectly valid.

Figure~\ref{fig:Sz}(a) demonstrates an example of the measured Kerr
rotation signal in the pump-probe delay range from $0$ to $13.2$\,ns
at $B=25.5$\,mT (black curve). To evaluate $S_z$, which is directly
proportional to the KR amplitude, we fit the data with an
exponentially decaying cosine function and integrate the fit curve
over the whole period of $T_\text{R}=13.2$\,ns. The magnetic field
dependence of $S_z$ is given in Fig.~\ref{fig:Sz}(b). It has a
finite value for $B<15$\,mT and approaches zero for higher fields.
It is instructive to compare the $S_z(B)$ dependence with the
results of the RSA shifts shown again in Fig.~\ref{fig:Sz}(c). From
this comparison one can conclude, that the high nuclear spin
polarization, corresponding to large shifts of RSA peaks, is induced
in magnetic fields, where $S_z$ is already zero.

\subsubsection{$S_x$: Knight field influence}\label{subsubsec:Sx}

\begin{figure}[t]
\includegraphics[width=\columnwidth]{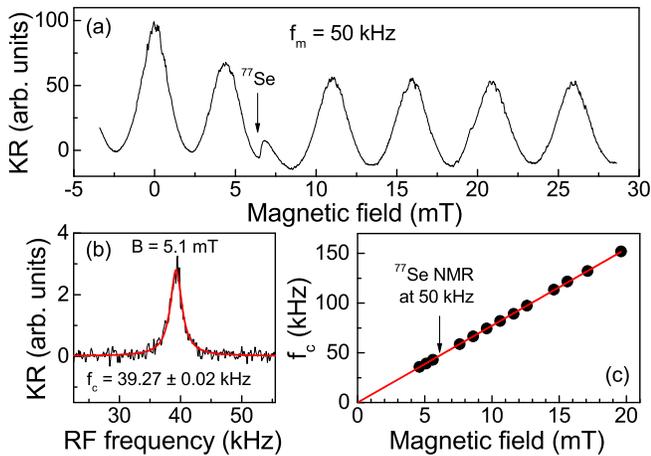}
\caption{(Color online) (a) RSA curve measured
for$f_\text{m}=50$\,kHz at $P=8$\,mW. (b) RF frequency scan at fixed
$B=5.1$\,mT. $f_c$ gives the central frequency defined by the
Lorentzian fit (red curve) to the data. (c) Dependence of the $f_c$
on the magnetic field. Red line is a linear fit to the data.}
\label{fig:Sx}
\end{figure}

The presence of an average electron spin polarization has an effect
on the nuclei through providing an effective magnetic field, the so
called Knight field, see Sec.~\ref{subsec:theory}~\cite{OptOR}. To
evaluate its value along the external magnetic field, which is
proportional to $S_x$ spin component, one can evaluate the effect of
the Knight field from the NMR frequency dependence on the magnetic
field. For this purpose, we have scanned the RF-field frequency for
$U_{\mathrm{RF}}= 0.05$\,V around the $^{77}$Se NMR for different
magnetic fields. Any Knight field component produced by an average
$S_x$ component would induce an additional magnetic field along the
$x$-axis and thereby lead to an offset of the linear dependence of
NMR resonance of the $^{77}$Se isotope on the magnetic field.
Figure~\ref{fig:Sx}(a) shows the RSA curve measured for
$f_\text{m}=50$\,kHz at $P=8$\,mW, where the $^{77}$Se NMR is seen
at about $B_{\text{NMR}}= 6$\,mT. Figure~\ref{fig:Sx}(b) gives an
example of the RF-frequency scan at a fixed magnetic field of
$B=5.1$\,mT. As shown by the red fit using a Lorentz curve, the
central NMR frequency is located at $f_c=39.37 \pm 0.02$\,kHz,
showing also the precision of the measurement of the NMR position:
$0.02$\,kHz accuracy corresponds to about $3$\,$\mu$T field
strength. Finally, Fig.~\ref{fig:Sx}(c) represents a collection of
all measured $f_c$ at different magnetic fields. This data set
demonstrates a linear dependence of the resonance frequency with
magnetic field where the corresponding fit leading to an offset of
$-0.4\pm0.2$\,kHz, which corresponds to $-51\pm 26$\,$\mu$T. This
allows us to conclude, that the Knight field produced by a possible
$S_x$ spin component, if present at all, should not be the reason
for the measured nuclear spin polarization. As will become clear
from the next sections, such an offset (or a Knight field) is
negligible compared to the Knight field produced along the $y$
direction. Namely, at a distance of one localization radius $a_l$
from the donor center, the Knight field reaches a strength of
$3.5$\,mT along $y$ direction, see Sec.~\ref{subsec:theoryspin}.

\subsubsection{$S_y$: RF-field versus Knight field}\label{subsubsec:Sy}

To test the presence of a $S_y$ spin component we applied a RF-field
with the same frequency as the used helicity modulation frequency
$f_\text{m}$. The relative phase between this RF-field and the
helicity modulation could be controlled, as well as the RF
amplitude.

\begin{figure*}[t]
\includegraphics[width=\textwidth]{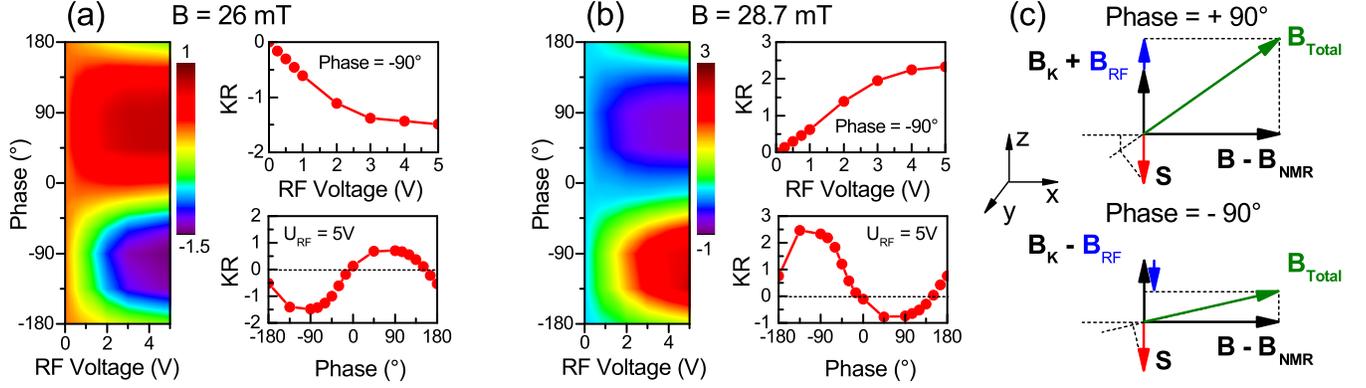}
\caption{(Color online) (a) Contour plot for the phase between the
RF field applied along the $z$ axis and the helicity modulation for
different RF amplitudes at a magnetic field strength of 26~mT. On
the righthand side of the plot cuts through the contour at the phase
of $-90^{\circ}$ (top) and at the RF voltage of 5\,V (bottom) are
shown. (b) Same as in (a) but for a different magnetic field
position, namely 28.7~mT. $f_\text{m}=185$\,kHz. (c) Schemes of the
rotating frame systems (RFS)  for the RF field acting along the
Knight field $+90^{\circ}$ and antiparallel to it $-90^{\circ}$.
Both schemes are given for $B>B_{\text{NMR}}$.} \label{fig:RFPhase}
\end{figure*}

In the frame system rotating with frequency $f_\text{m}$, one can
interpret the RF-field as an additional, constant-in-time magnetic
field acting on the nuclei (see Sec.~\ref{subsec:theory} on the
rotating frame system). This allow us to find the orientation of the
average electron spin component and compensate its action on the
nuclei by counteracting with the applied RF-field, see
Fig.~\ref{fig:RFPhase}(c).

Figure~\ref{fig:RFPhase} demonstrates the results of such
measurements at different magnetic fields for $f_\text{m}=185$\,kHz.
The magnetic field values are comparable to those applied in the RSA
studies Fig.~\ref{fig:main}(a). Here we plot the signal determined
as the difference between the KR amplitudes with and without the RF
field, applied at a fixed magnetic field for a high pump power of
$P=8$\,mW. Figure~\ref{fig:RFPhase}(a) shows the influence of the RF
field on the KR amplitude at the field position of $B=26$\,mT on the
right hand side of a RSA peak, i.e. in a field range where the KR
signal has a decreasing slope with increasing field. The color
coding is such that blue color means reduction of the Kerr rotation
signal under the influence of the applied RF-field, while red color
represents an increase of the signal. The panels right next to the
contour plot show cuts through the contour plot, namely the RF
amplitude dependence at fixed phase of $-90^{\circ}$ (top) and the
phase dependence at the saturated RF amplitude level of
$U_{\mathrm{RF}}=5$\,V, which corresponds to an effective magnetic
field of $170$\,$\mu$T (bottom)~\footnote{The calibration of the RF
coil for conversion from applied voltage to magnetic field is done
using the Rabi rotation of the $^{77}$Se isotope. For our RF coil
the following relation holds: $B/U_{\mathrm{RF}} = 34$\,$\mu$T/V.
These measurements will be discussed elsewhere.}.

As one can see from the RF dependence, the Kerr amplitude is reduced
for higher RF voltages, i.e. the RSA peak shifts to lower magnetic
fields so that the nuclear spin polarization is reduced. This is
most efficient for the phase close to $-90^{\circ}$, implying that
the Knight field component should initially be oriented along the
$y$ axis. In the transverse magnetic field it becomes then rotated
to the direction parallel to the $z$ axis, thereby accumulating the
phase of $-90^{\circ}$, which is best compensated when the RF field,
applied along the $-z$ axis, is acting directly against it, see
Fig.~\ref{fig:RFPhase}(c). As an additional support for this
interpretation, we observe an increase of the KR amplitude at
$+90^{\circ}$, where the nuclear spin polarization becomes amplified
by the RF field. Here the Knight field is acting in the same
direction as the RF, see Fig.~\ref{fig:RFPhase}(c), increasing
therefore the overall nuclear spin projection $I_{\text{N},x}$ along
the $x$-axis.

A similar behavior is observed at a higher magnetic field strength,
$B=28.7$\,mT, as shown in Fig.~\ref{fig:RFPhase}(b). Here we are on
the left increasing slope of the next RSA peak, so that the overall
amplitude changes are inverted. The RSA peak shifts to lower
magnetic fields for reduced nuclear field and the KR amplitude
increases. The relative phase behavior of the RF however stays the
same: we reduce the nuclear spin polarization at $-90^{\circ}$ and
increase it at $+90^{\circ}$, which shows that the Knight field has
a constant phase shift relative to the RF field and is oriented
along the $y$ axis.

Several more RSA peaks at higher magnetic fields were tested
demonstrating similar behaviors (not shown here). Measurements at
different magnetic fields demonstrate that the electron spin
orientation stays the same. Additionally, the effect of the RF field
at the phase of $-90^{\circ}$ leads to similar changes of the KR
amplitude in the range of fields $20-40$\,mT which has an value of
about $1.9 \pm 0.3$ (arb. units of KR amplitude). This means, that
there is a finite average electron spin component along the $S_y$
direction, which stays constant when varying the transverse magnetic
field strength.

\section{Theoretical model and discussion}

\subsection{Classical treatment}\label{subsec:theory}

\begin{figure}[t]
\includegraphics[width=\columnwidth]{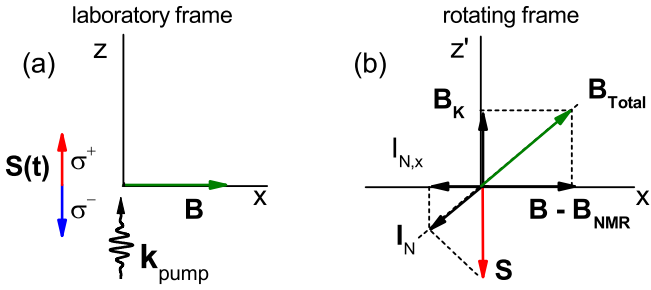}
\caption{(Color online) (a) Spin orientation in the laboratory frame
system. The red and blue arrows symbolize the electron spins
generated with $\sigma^+$ and $\sigma^-$ photons, respectively,
alternating in helicity with $f_{\text{m}}$. (b) Scheme of magnetic
fields, acting on the nuclei in the rotating frame. The frame is
rotating about the \textit{x} axis with frequency $2 \pi
f_{\text{m}}$ in the direction of the nuclear spin precession.
$\mathbf{B}_{\text{K}}$ and $\mathbf{S}$ are the average electron
Knight field and spin, respectively, which are fixed in the rotating
frame and are rotating with $2 \pi f_{\text{m}}$ in the laboratory
frame. $I_{\text{N},x}$ is the projection of the induced nuclear
spin polarization onto the direction of the external field
$\mathbf{B}$.} \label{fig:systems}
\end{figure}

Let us consider excitation with circular polarization alternated
between left and right at the modulation frequency $f_\text{m}$. An
external magnetic field is applied perpendicular to the pump light
\emph{k} vector, $\textbf{k}_{\text{pump}}$, which is shine in
parallel to the \emph{z} axis. Figure~\ref{fig:systems}(a) shows
this configuration in the laboratory frame system (LFS). It is known
that under helicity modulated excitation with modulation frequency
$2\pi f_\text{m}\gg 1/T_1^{\text{nucl}}$ an optical polarization of
the nuclear spins occurs only for $B$ close to the resonance field
$B_{\text{NMR}}=2\pi f_\text{m}/\gamma$. Here $T_1^{\text{nucl}}$ is
the nuclear spin relaxation time and $\gamma$ is the gyromagnetic
ratio of the nuclear isotope~\cite{OptOR,Kalevich1980}. Measurements
of $T_1^{\text{nucl}}$ will be presented elsewhere. For the studied
ZnSe:F epilayer these times fall in the range of tens of
milliseconds. The literature values of $\gamma$ for the $^{77}$Se
($\gamma_{Se}=51.08\times 10^6$\,[T s]$^{-1}$) and $^{67}$Zn
isotopes ($\gamma_{Zn}=16.77 \times 10^6$\,[T s]$^{-1}$) lead to
$B_{\text{NMR}}$\,[mT]$ = 0.123 f_\text{m}$\,[kHz] and $0.375
f_\text{m}$\,[kHz], respectively~\cite{PhysHandbook}. These values
are in very good agreement with the values observed for the NMR
resonances.

In the following we consider only the $^{77}$Se isotope having
nuclear spin $I=1/2$ with an abundance $\chi=0.0758$. The hyperfine
constant $A_{Se}=33.6$\,$\mu$eV, which is taken for a primitive cell
with two nuclei~\cite{Syp11,Syp11E}.

The dynamic nuclear spin polarization is caused by nuclear spin
flips in the presence of the electron Knight field
$\mathbf{B}_{\text{K}}=b_{\text{e}} \mathbf{S}$ which precesses
synchronously with the electron spin $\mathbf{S}$ and provides a
temporally constant energy flow into the nuclear spin
system~\cite{OptOR_Flei_Merk}. Here
\begin{equation}
\label{eq:be}
b_{\text{e}}=-\frac{A_{Se}v_0}{\gamma_{Se}\hbar\pi a_l^3},
\end{equation}
characterizes the maximal Knight field amplitude at the center of
the donor~\cite{OptOR_Flei_Merk}. $v_0=a_0^3/4$ is the primitive
cell volume with a two-atom basis and $a_0=0.566$\,nm is the lattice
constant of ZnSe~\cite{Paw11}; $a_l$ gives the localization radius
of an electron at the donor.

The average nuclear spin polarization $\mathbf{I}_{\text{N}}$ has a
component along the external field $\mathbf{B}$, $I_{\text{N},x}$,
and, therefore, affects the electron spin precession frequency, see
Fig.~\ref{fig:systems}(b). If the external magnetic field is close
to the $B_{\text{NMR}}$, the projection of the nuclear field on the
external field $\mathbf{B}$ is given by:
\begin{equation}
\label{eq:1}
I_{\text{N},x}=\frac{({\mathbf{S}}\cdot\mathbf{B}_{\text{K}})(B-B_{\text{NMR}})}{(B-B_{\text{NMR}})^2+B^{2}_{\text{K}}+B^{2}_{\text{L}}},
\end{equation}
where $B_{\text{L}}=0.006$\,mT is the root mean square local field
due to the nuclear dipole-dipole interactions~\cite{OptOR,Zhu2014}.

This can be interpreted as nuclear spin cooling in the rotating
frame system (RFS)~\cite{Kalevich1980}, see
Fig.~\ref{fig:systems}(b). The Knight field, oscillating with
frequency $2\pi f_{\text{m}}$, can be described by a superposition
of two fields, each rotating around the external field opposite to
each other. Close to the NMR frequency the component, which rotates
in the same direction as the nuclear spin is the important one,
while the other one can be neglected. In ZnSe all nuclei have
$\gamma>0$~\cite{PhysHandbook}, therefore this component is rotating
counterclockwise if one looks in the direction along the external
field $\textbf{B}$. In the RFS the electron spin is constant and the
nuclei see the total magnetic field $\mathbf{B}_{\text{Total}}$,
composed of the constant Knight field $\mathbf{B}_{\text{K}}$ and
the effective external field $\mathbf{B}-\mathbf{B}_{\text{NMR}}$.
This situation is analogous to the cooling in the laboratory frame
system (LFS) in the stationary case. One can see from
Eq.~\eqref{eq:1} that upon fulfilling the resonance condition
$B=B_{\text{NMR}}$ the nuclear spin polarization along the external
magnetic field $I_{\text{N},x}=0$.

The polarized nuclei create an Overhauser field (see
Ref.~\cite{Overhauser1953}) which acts on the electron spins:
\begin{equation}
\label{eq:Bnx}
B_{\text{N},x}=\frac{A_{Se} {\chi} I_{\text{N},x}}{\mu_\mathrm{B} g_{\text{e}}}.
\end{equation}
Here $\mu_\mathrm{B}$ is the Bohr magneton. From Eq.~\eqref{eq:1}
one can see that the $B_{\text{N},x}(B)$, following
$I_{\text{N},x}(B)$, has a dispersive shape. Its sign is defined by
the detuning $B-B_{\text{NMR}}$. This describes exactly the behavior
of the signal described in Sec.~\ref{subsec:experiment}, item (i).

By adding to or subtracting from the external magnetic field, the
field $B_{\text{N},x}$ alters the electron spin precession
frequency. In turn, this leads to a change of the electron spin
polarization value at a specific external magnetic field. Usually,
this effect occurs if the resonant field $B_{\text{NMR}}$ does not
exceed the half width of the electron spin depolarization curve
$B_{1/2}$ (Hanle curve)~\cite{Kalevich1980,Cherbunin11}. Otherwise,
the electrons become depolarized due to their spin precession about
the transverse field $\mathbf{B}+\mathbf{B}_{\text{N},x}$, so that
the value $(\textbf{S}\cdot \textbf{B}_{\text{K}})$ and the
resulting $B_{\text{N},x}$ are small. Nevertheless, it has been
shown in Ref.~\cite{Zhu2014} that such resonances do occur also at
the $B_{\text{NMR}}$ fields, which are much stronger than $B_{1/2}$
and have a width of about one mT, see item (ii). In what follows, we
concentrate on the nature of the broad resonance (item (iii)).

As one can see, Eqs.~\eqref{eq:1} and~\eqref{eq:Bnx} support the
measurements presented in Figs.~\ref{fig:SP}(a) and \ref{fig:SP}(b).
The nuclear spin polarization, which causes the RSA shift via the
Overhauser field $B_{\mathrm{N},x}$, indeed follows the $S^2$
dependence. The $S^2$ dependence comes from the product
$(\textbf{S}\cdot \textbf{B}_{\text{K}})$, where
$\mathbf{B}_{\text{K}}=b_{\text{e}} \mathbf{S}$. The accelerated
decay of the nuclear spin polarization with increased magnetic field
is also well described by the $S^2$ dependence in the denominator of
Eq.~\eqref{eq:1}, see Fig.~\ref{fig:SP}(c).

As mentioned before, the sign of the induced shifts in
Fig.~\ref{fig:main}(b) is opposite to the induced nuclear fields,
$B_{\text{N},x}$. Taking into account that in ZnSe the electron $g$
factor $g_{e}>0$, so that $b_e<0$ and that the hyperfine constant
$A_{Se}>0$, Eq.~\eqref{eq:Bnx} reproduces this sign dependence. It
is negative for $B>B_{\text{NMR}}$ and positive vice versa.

\begin{figure}[t]
\includegraphics[width=\columnwidth]{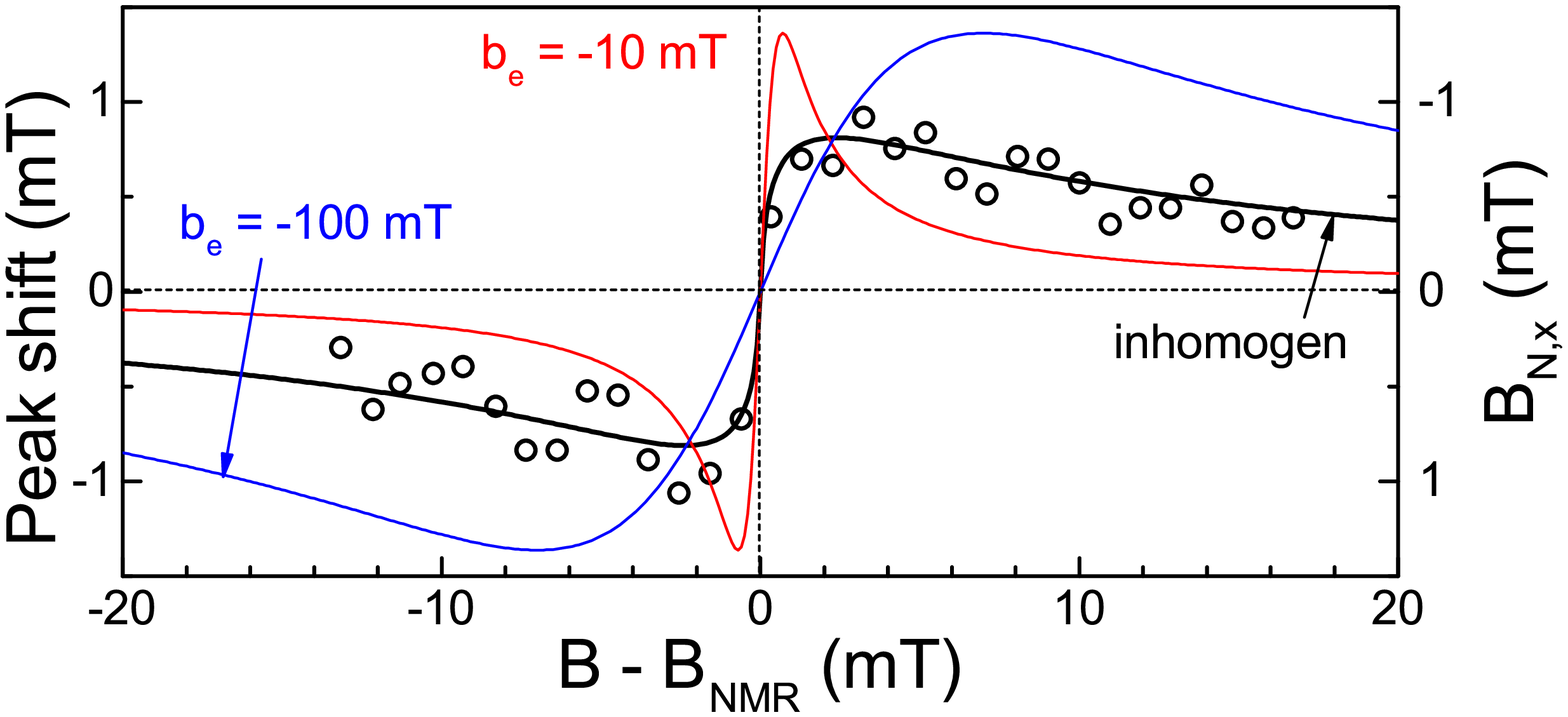}
\caption{(Color online) Peak shift and corresponding
$B_{\text{N},x}$ calculated for the $^{77}$Se isotope after
Eq.~\eqref{eq:3} (black line) with $b_{\text{e}} = -370$\,mT,
$S=0.07$ and $a_l=3.4$\,nm. For comparison we show two homogeneously
broadened curves calculated after Eq.~\eqref{eq:Bnx} using
$b_{\text{e}} = -10$\,mT (red line) and $b_{\text{e}} = -100$\,mT
(blue line). Open circles represent the experimental data as in
Fig.~\ref{fig:main}(b), but shifted slightly downwards to compensate
for the offset.} \label{fig:inhomo}
\end{figure}

We now try to simulate on the basis of Eq.~(\ref{eq:Bnx}) the data
shown by the open circles in Fig.~\ref{fig:main}(b) and reproduced
in Fig.~\ref{fig:inhomo}. The fitting parameters in this procedure
are $b_e$ and $S$. $S$ is fixed at $0.07$, which will be justified
below. Figure~\ref{fig:inhomo} demonstrates examples for a
homogeneous nuclear spin polarization using $b_e=-10$\,mT and
$-100$\,mT, given by the red and blue curves, respectively. As can
be seen, the $b_e=-10$\,mT curve describes only the fast changeover
of the sign close to the resonance very well, but completely fails
to fit the data away from the resonance. On the other hand, the
$b_e=-100$\,mT curve shows the right tendency compared to the data
only in the tails, far from the resonance condition. The amplitudes
here, however, deviate considerably from the measured data points.

This simulation leads to an important conclusion: the Knight field
is not constant within the localization area of the donor-bound
electron. As is well known, a uniform spin polarization of the
nuclei can be established through spin diffusion based on flip-flop
processes between the nuclei at different distances relative to the
donor center~\cite{DyakonovBasics,Bloembergen1954}. Spin diffusion
is allowed, if the energy is conserved (or nearly conserved)
thereby, so that the energy difference between spin flips of two
nuclei does not exceed $\hbar \gamma B_{\text{L}}$, and therefore,
can be compensated by the dipole-dipole nuclear reservoir. In our
case, however, the nuclei are exposed to an inhomogeneous Knight
field $B_{\text{K}}=b_{\text{e}}(r)S=b_{\text{e}} S
\exp{(-2r/a_l)}$, which is different for neighboring nuclei and is
given by the electron wave function at the donor: $\Psi^2(r)=(\pi
a^3_l)^{-1} \exp(-2r/a_l)$. Here $r$ is the distance from the donor
center. Neighboring nuclei of $^{77}$Se are separated by a distance
of about $R=a_0/\chi_{\text{Se}}^{1/3}=
0.566/0.0758^{1/3}=1.34$\,nm~\cite{Greilich12}. The difference of
the Knight field at neighboring isotopes is than in the order of
$b_{\text{e}} S \exp{(-2r/a_l)}[1-\exp{(-2R/a_l)}]$. This should be
compared with the local nuclear field $B_{\text{L}}\approx
0.006$\,mT. Therefore the radial diffusion of the nuclear spin
becomes only possible if $b_{\text{e}} S
\exp(-2r/a_l)[1-\exp(-2R/a_l)] \leq B_{\text{L}}$. To estimate the
$r$ at which the nuclear spin diffusion becomes possible, we need to
know the values for $b_e$ and $S$.

The activation energy (or donor binding energy) of the electron
bound to the fluorine donor ($E_{\text{a}}=27$\,meV) allows us to
estimate the localization radius of the electron $a_l=3.4$\,nm using
$a_l=\hbar / \sqrt{2 m^{\text{eff}}_{\text{e}}} E_{\text{a}}$, with
$m^{\text{eff}}_{\text{e}}=0.145 m_{\text{e}}$~\cite{Greilich12},
where $m_{\text{e}}$ is the free electron mass in vacuum. This small
$a_l$ value should lead to a significant Knight field at the donor
center. Using Eq.~\eqref{eq:be} given at the beginning of the
Sec.~\ref{subsec:theory} we obtain $b_{\text{e}}=-370$\,mT~\footnote{The $a_l$ (and corresponding $b_{\text{e}}$) is different here in comparison to Ref.~\cite{Zhu2014}. The value of $a_l=7$\,nm in Ref.~\cite{Zhu2014} was taken to provide a smaller Knight field for a reasonable fitting to the NMR appearance in the RSA curve. The Knight field was considered as homogeneous.}. Then,
taking into account that $S=0.07$ (see below in this section) this
leads to the result, that the nuclear spin diffusion should be
hindered within a radius of about $3.9$\,$a_l$, resulting in this
range in a spatially inhomogeneous nuclear spin polarization,
$I_{\text{N},x}(r)$.

Due to the spherical symmetry of the spin diffusion the polarization
is also isotropic. Thus, the spatial distribution of the nuclear
spin polarization is given by:
\begin{equation}
\label{eq:2}
I_{\text{N},x}(r)=\frac{b_{\text{e}} S^2
(B-B_{\text{NMR}}) \exp{(-2r/a_{\text{l}})}}{(B-B_{\text{NMR}})^2+
b_{\text{e}}^2 S^2 \exp{(-4r/a_{\text{l}})}},
\end{equation}
We neglect here the small $B_\mathrm{L}$ fields. The polarized
nuclei, in turn, act on the electrons via the Overhauser field:
\begin{equation}
\label{eq:3}
B_{\text{N},x}=\frac{A{\chi}}{\mu_{\text{B}} g_{\text{e}}}\int I_{\text{N},x}(r) \Psi^2(r) 4\pi r^2 dr.
\end{equation}

We use Eq.~\eqref{eq:3} to fit the experimental data for the induced
nuclear field given in Fig.~\ref{fig:inhomo}. The best fit is shown
by the black curve. It has been achieved for $b_{\text{e}}=-370$\,mT
and an average electron spin polarization of $S=0.07$, and
reproduces all features of the experimental dependence. Namely, the
fast changeover of the field sign at the resonance
$B_{\mathrm{NMR}}$ and the broad tails due to the widely extended
decay.

\begin{figure}[t]
\includegraphics[width=\columnwidth]{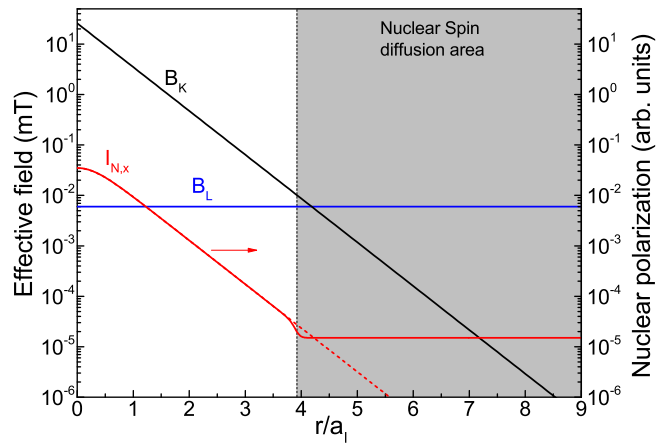}
\caption{(Color online) Illustration of different fields as function
of the distance from the donor center in units of the localization
radius $a_l$. $B_{\text{K}}$, black line, is the Knight field. Here
the simulation is done using $b_{\text{e}} S = 25.9$\,mT. Grey
shaded area shows the region where the spin diffusion in the nuclear
spin system becomes possible. $I_{\text{N},x}$, red dashed line,
shows the nuclear spin polarization using $B-B_{\text{NMR}} = b_e
S$. Red solid line represents schematically the effect of nuclear
spin diffusion on $I_{\text{N},x}$, which equalizes the nuclear spin
polarization spatially for $r > 3.9$\,$a_l$. Blue line shows the
strength of the local fields $B_\mathrm{L}=0.006$\,mT.}
\label{fig:Cake}
\end{figure}

Figure~\ref{fig:Cake} shows schematically the behavior of the Knight
field ($B_{\text{K}}=b_{\text{e}} S \exp{(-2r/a_l)}$) and the
nuclear spin polarization ($I_{\text{N},x}$) using Eq.~\eqref{eq:2}
as a function of the distance from the donor. As soon as the radius
becomes larger than $3.9$\,$a_l$ spin diffusion becomes possible and
equalizes the nuclear spin polarization spatially, as shown
schematically by the red solid line showing the region of flat
nuclear spin polarization above $3.9$\,$a_l$.

Using Figs.~\ref{fig:inhomo} and~\ref{fig:Cake} one can draw the
following qualitative conclusions: (a) close to the resonance
($B=B_{\text{NMR}}$) the electron is exposed to a very weak nuclear
spin polarization produced by the Knight field far from the donor,
at the edges of the nuclear spin diffusion area. As the
concentration of fluorine donors in this sample is $n=1\times
10^{18}$\,cm$^{-3}$, the average distance between them is
$\bar{d}=\left(\frac{3}{4\pi n}\right)^{1/3}=62$\,nm. This
corresponds to about $18 a_l$, so that the donors are not located in
the inhomogeneous nuclei polarisation volumes of the neighbors. (b)
At the center of the donor, the electrons are exposed to the
strongest nuclear spin polarization produced by the maximal Knight
field. This is the position far from the resonance on the magnetic
field axis in the extended tails of our inhomogeneous dispersive
curve in Fig.~\ref{fig:inhomo}. A simple estimation using the
integral $\int \Psi^2(r) 4\pi r^2 dr$ shows, that the electron is
98\% confined within the volume given by the radius $3.9$\,$a_l$.
The nuclear spin polarization decreases at that distance from the
donor by a factor of about 1200. If the spin diffusion border would
be much closer to the donor or, with other words, if $B_{\text{L}}$
would be bigger, the extended tails would decay much faster with
magnetic field. For example, in GaAs $B_{\text{L}}=0.3$\,mT~\cite{OptOR_Flei_Merk}. If
taking into account $b_e\approx 20$\,mT as well as
$a_l=10$\,nm~\cite{OptOR_Flei_Merk} and using similar considerations
with $S=0.07$ as for ZnSe, one can estimate the diffusion border to
be at $r \approx 0.3 a_l$. This small borderline would give a quite
homogenous nuclear spin polarization around the donor and lead to a
narrow NMR resonance, experimental examples are given in
Ref.~\cite{Zhu2014}.

\subsection{Average electron spin polarization in external magnetic field}\label{subsec:theoryspin}

The next point to clarify is the value of the average electron spin
polarization in the range of magnetic fields around the NMR field.
These fields are much higher than the half width of the Hanle curve
($B_{1/2}$). In our simulations using Eq.~\eqref{eq:3} we have
estimated the average electron spin value to be constant at
$S=0.07$. The value of the average nuclear spin polarization given
in Eq.~\eqref{eq:1} is proportional to
$({\mathbf{S}}\cdot{\mathbf{B}_{\text{K}}})=b_{\text{e}} S^2(t=0)
B_{1/2}^2/(B_{1/2}^2+B^2)$ and is expected to be reduced by a factor
of $(B/B_{1/2})^2 \gg 1$ for higher magnetic fields~\cite{OptOR}.
Here $(B/B_{1/2})^2 \approx (20$\,[mT]/$1$\,[mT]$)^2 = 400$ and the
average electron spin transverse to the magnetic field should be
very small due to the Larmor precession of the electron spin about
this field. This statement agrees with our experimental observations
presented in Fig.~\ref{fig:Sz}(b) in Sec.~\ref{subsec:spinpol}.

\begin{figure}[t]
\includegraphics[width=\columnwidth]{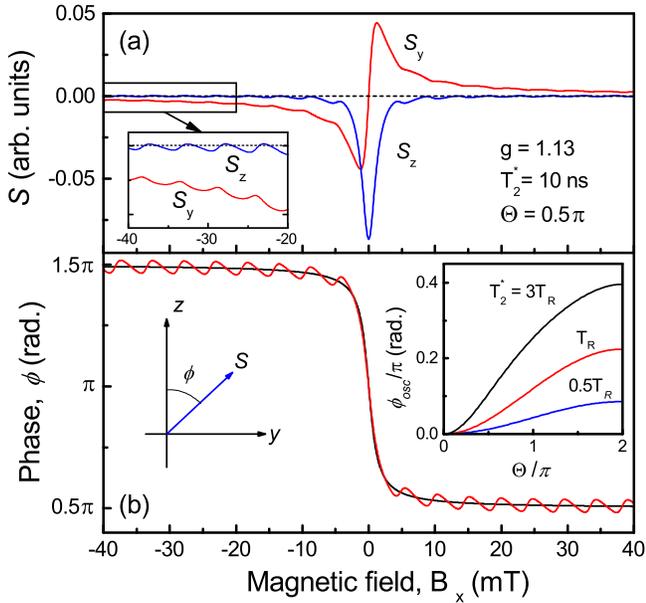}
\caption{(Color online) (a) Evolution of the average spin components
in transverse magnetic field $\mathbf{B}=(B_x,0,0)$. Inset shows a
close-up for magnetic fields from -40 to -20\,mT to demonstrate the
relative weights of the spin components. (b) Phase relation between
the spin components. The amplitude of the oscillations in the red
curve can be related to the red curve in the inset. This amplitude
depends on the spin dephasing in relation to the laser repetition
period, over which the averaging is done. The longer the spin
dephasing takes, the smaller the amplitude that is expected for the
oscillations, see black curve in the inset and panel (b).}
\label{fig:spin}
\end{figure}

However, the experimental observation of the broad resonance allows
us to assume that there is an average electron spin polarization
present in a wide range of magnetic fields, which leads to
polarization of the nuclei, in particular along the external
magnetic field axis, seen as $I_{\text{N},x}$ in
Fig.~\ref{fig:systems}(b). Using Eq.~(13) of Ref.~\cite{Zhu2014} we
can estimate the average electron spin polarization along the $y$
and $z$ directions, where $z$ is the optical excitation axis. The
averaging is done over the period of the laser repetition,
$T_{\text{R}}$. Figure~\ref{fig:spin}(a) demonstrates the evolution
of the average electron spin components with the transverse magnetic
field, $B_x$. The $S_z$ component decays completely within a range
of 10\,mT (which fits well to our observations in
Fig.~\ref{fig:Sz}), while the $S_y$ component is slowly decaying up
to tens of mT, see the inset to panel (a) magnifying the difference
at higher fields. As mentioned in Ref.~\cite{Zhu2014}, the vector
sum of the $z$ and $y$ spin components decays with increasing
transverse field $B$ as $1/B$.

Figure~\ref{fig:spin}(b) shows the corresponding evolution of the
phase $\phi$ between the $S_z$ and $S_y$ spin components. It allows
us to conclude, that the average spin polarization along the $y$
direction should be present and have a constant phase shift of
$90^{\circ}$ relative to the $S_z$ component in the range of
magnetic fields above 5\,mT. The inset demonstrates the power
dependence of the phase-oscillation amplitude present in the
simulated signal for different electron spin dephasing times,
$T_2^*$, in relation to the laser repetition period $T_{\text{R}}$.
The red curve represents the simulation using the realistic
parameters of our system, $T_2^*=10$\,ns and
$T_{\text{R}}=13.2$\,ns. $\Theta$ is the optical pulse area,
$\Theta= \int 2\langle d \rangle E(t)dt / \hbar$, where $\langle d
\rangle$ is the dipole transition matrix element and $E(t)$ is the
electric field of the laser pulse~\cite{Zhu2014}. $\Theta=\pi$
corresponds to the power for  100\% exciton generation.

Therefore, the only spin component which can potentially provide the
link to the nuclear spin polarization at high magnetic fields is the
$S_y$ average spin component. However, even if the average electron
spin decays as $1/B$ the nuclear spin polarization is proportional
to $S^2$ (as seen shown in Sec.~\ref{subsec:spinpol}) and should
therefore decay as $1/B^2$. This should suppress the nuclear spin
polarization drastically with increasing magnetic field.

On the other hand the experiments with RF field described in
Sec.~\ref{subsubsec:Sy} demonstrate, that the $S_y$ spin
polarization does not decay within the measured magnetic field range
$B=20-40$\,mT. Also, as one can see in Fig.~\ref{fig:mod}(b), the
amplitudes of the induced shift do not depend on the NMR position in
the magnetic fields up to 140\,mT~\footnote{We are not taking into
account here the overall vertical shift of the curves.}. These two
observations lead us to the surprising result, that the average
electron spin polarization $S_y$ does not change in the measured
field range.

The presented classical model gives us the possibility to describe
the overall behavior of the signal with all its peculiarities.
However, it requires presence of a relatively strong electron spin
polarization $S_y$ that does not change with magnetic field. Such a
polarization could be caused by certain anisotropies intrinsic to
ZnSe:F. First, the fluorine atom itself being placed at the position
of selenium could lead to such an anisotropy. Second, another
possibility could be an anisotropy of the electron spin generation
provided by strain in the crystal lattice. Both these assumptions
require further investigations and will be presented elsewhere.

Further possibilities to generate DNP close to RSA peaks may be also
considered. For example, for sufficiently high pump power, an
effective magnetic field along the $x$ axis could be induced by the
interaction of the absorption resonance with circularly polarized
light (optical Stark effect). In that case pulsed excitation, where
laser pulses hit the sample in phase with the electron spin
$\mathbf{S}(t)$ at multiple frequencies of the laser repetition
period $T_R$, induces a $S_x$ electron spin polarization along the
external magnetic field~\cite{Kor2011}. However, the sign of the
induced spin polarization should depend on the relative energy
between the absorbtion and the excitation energy. This possibility
was excluded by measuring the induced nuclear field as a function of
the optical excitation energy. The induced shift had no sign changes
for excitation below and above the D$^0$X resonance.

\section{Conclusions}\label{sec:conclusion}

We have considered here that the shift of the electron Larmor
frequencies or peaks in RSA signal in ZnSe:F under high excitation
power is result of a dynamic nuclear spin polarization. This
statement is confirmed by the position of the NMR resonance and its
dependence on the modulation frequency $f_{\text{m}}$.

The shape and the width of the resonance allow us to conclude that
it should be driven by the inhomogeneous Knight field acting on the
nuclear system. This Knight field has a weak dependence on external
magnetic field and is pointing along the $y$ direction. This
assumption is confirmed by measurements, where an additional RF
field is used to compensate the effect of the Knight field. The
estimated values of the Knight field lead to the conclusion that the
nuclear spin diffusion is hindered within a radius of about
$3.9$\,$a_l$ from the fluorine donor center leading in this range to
inhomogeneous nuclear fields, which in turn contribute to the broad
NMR resonance seen in experiment.

\begin{acknowledgments}
We acknowledge the financial support by the Deutsche
Forschungsgemeinschaft in the frame of the ICRC TRR 160, the
Volkswagen Stiftung (Project No.\,88360/90080) and the Russian Science
Foundation (Grant No. 14-42-00015). T.K. acknowledges financial
support of the Project “SPANGL4Q” of the Future and Emerging
Technologies (FET) programme within the Seventh Framework Programme
for Research of the European Commission, under FET-Open Grant No.
FP7-284743. V.L.K. acknowledges financial support from the Deutsche
Forschungsgemeinschaft within the Gerhard Mercator professorship
program.
\end{acknowledgments}

\bibliographystyle{apsrev}
\bibliography{ZnSeF_shift_refs}

\end{document}